\documentclass[a4paper,fleqn,usenatbib]{mnras}

\usepackage{newtxtext,newtxmath}
\usepackage[T1]{fontenc}
\usepackage{ae,aecompl}

\usepackage{graphicx}	% Including figure files
\usepackage{amsmath}	% Advanced maths commands
\usepackage{amssymb}	% Extra maths symbols

\usepackage{rotating}
\usepackage{threeparttable}

\usepackage{dcolumn}
\usepackage{longtable}
\usepackage{multirow}
\usepackage{bm}
\usepackage{textpos}

\title[High Velocity Runaway Binaries from Tertiary Supernovae in Triple Stellar Systems]{High Velocity Runaway Binaries from Supernovae in Triple Systems}

\author[Y. Gao et al.]{
Yan Gao,$^{1,2,3}$\thanks{E-mail: ygbcyy@ynao.ac.cn}
Jiao Li,$^{1,2,3}$
Shi Jia,$^{1,2,3}$
\\
$^{1}$Yunnan Observatories, Chinese Academy of Sciences, Kunming 650011, China\\
$^{2}$Key Laboratory for the Structure and Evolution of Celestial Objects, Chinese Academy of Sciences, Kunming 650011, China\\
$^{3}$University of Chinese Academy of Sciences\\
}

\date{Accepted XXX. Received YYY; in original form ZZZ}

\pubyear{2019}

\begin{document}
\label{firstpage}
\pagerange{\pageref{firstpage}--\pageref{lastpage}}
\maketitle

% Abstract of the paper
\begin{abstract}
Recent studies on hypervelocity stars (HVSs) have generated a need to understand the high velocity limits of binary systems. If runaway binary systems with high movement speeds well in excess of 200km/s were to exist, it would have implications on how HVS candidates are selected, and our current understanding of how they form needs to be reinforced. In this paper, we explore the possibility that such high velocity runaway binaries (HVRBs) can be engendered by supernova explosions of the tertiary in close hierarchical triple systems. We find that such explosions can lead to significant remnant binary velocities, and demonstrate via constraining the velocity distribution of such HVRBs that this mechanism can lead to binaries with centre of mass velocities of 350 km/s or more, relative to the original centre of mass of the progenitor triple system. This translates into potential observations of binaries with velocities high enough to escape the Galaxy, once the Galactic rotational velocity and objects of Large Magellanic Cloud origins are considered. 
\end{abstract}

% Select between one and six entries from the list of approved keywords.
% Don't make up new ones.

%%%%%%%%%%%%%%%%%%%%%%%%%%%%%%%%%%%%%%%%%%%%%%%%%%

%%%%%%%%%%%%%%%%% BODY OF PAPER %%%%%%%%%%%%%%%%%%

\section{Introduction}

The recent release of Gaia DR2 \citep{2018A&A...616A...1G} has enhanced, among other things, our understanding of hypervelocity stars (HVSs), stars which have velocities in the Galactic reference frame that are high enough to achieve escape velocity. The breakthroughs were mainly due to spectroscopic radial velocities archived in the catalogue, which led to the discovery and identification of many candidates for such objects that were previously unknown \citep[e.g.][]{2018ApJ...868...25B,2018ApJ...866..121H,2018A&A...620A..48I,2018MNRAS.tmp.2466M,2018ApJ...865...15S}. Further discussion of the advances consequently made in the field of HVS studies is beyond the scope of this paper, and will therfore be omitted here.

While performing these studies, some authors have considered the possibility that the HVS candidates they were processing are actually unresolved binaries masquerading as single stars. If this were the case, binary orbital motion would contaminate the catalogued radial velocities, leading to noise in the final derived velocities. Noting this potential source of error, alongside many others, the authors adjusted their candidate selection criteria accordingly \citep[e.g.][]{2018MNRAS.tmp.2466M}. Some of these criteria could have been further optimised if the velocity distribution of runaway binaries were known. For instance, consider a hypothetical object travelling at a very high velocity, for which binarity cannot be ruled out by traditional methods. If we were to assume that it is, in fact, part of a binary, we could constrain the centre-of-mass velocity for the binary system, by accounting for any potential contamination due to binary orbital motion. If this constraint leads to the velocity lying beyond the velocity range expected for binary systems, then the possibility of a binary companion could be disregarded for this particular object. However, if insufficient information is available on the expected velocities of binary systems, which is currently the case, then it would be prudent to eliminate this object from the sample, which is what would have happened in all studies to date.

Refining data processing techniques aside, it certainly seems legitimate to speculate that some current or future HVS candidates could be binaries in disguise. \cite{2018ApJ...857..114W}, who studied the issue of binary and triple contamination in the Gaia DR1 general stellar sample, found that many binaries have indeed been misidentified as single stars, forming a ``binary main sequence" roughly 0.75 magnitudes above the main sequence when plotted in an HR diagram. Upon visual inspection of HVS candidates, it is found that several lie on this ``binary main sequence", including but not necessarily limited to {\it Gaia} DR2 1540013339194597376, which can be identified to be a HVS candidate via its proper motion of 900 km/s alone \citep{2018ApJ...868...25B}, and {\it Gaia} DR2 1976597452042003072, with a proper motion of 500 km/s. The list would have gone on, if not for recent issues with Gaia DR2 radial velocities \citep{2019MNRAS.tmp..280B}. The natural question to ask, then, would be whether it is possible that a binary can be accelerated to such velocities.

For single stars, numerous theories have been presented in the past on how high velocities can be attained. Binary systems interacting with the Galactic central black hole can have one of its stellar components consumed by the latter, ejecting the remaining component at velocities sufficient to become unbound \citep[e.g.][]{1988Natur.331..687H,2003ApJ...599.1129Y}; supernovae occurring in close binaries can impart to their companions velocities of hundreds of km/s \citep[e.g.][]{1961BAN....15..265B,2009A&A...508L..27W}, some of which exceed escape velocity, depending on their position in the Galactic potential; dynamical ejection from multiple systems can also lead to extremely high velocities, especially for low-mass stars in globular clusters \citep[e.g.][]{2009MNRAS.396..570G}. Yet of all these mechanisms, none can explain a hypothetical binary system travelling at, say, 200 km/s relative to the background Galactic orbital motion - the first two require the destruction of one component of a binary system, whereas the third can rarely eject intact binaries at even moderately high velocities, with the number of binary systems at a given velocity having a sharp cutoff at ${\sim}$150 km/s \citep{2012ApJ...751..133P}. It follows that the production of binaries travelling at higher velocities requires triple systems, and indeed it was found that binaries travelling at such velocities consisting of a main-sequence star and a neutron star can arise in small quantities from hierarchical triples where the tertiary is a supermassive black hole \citep{2019MNRAS.484.1506L}. It should also be easy to imagine that, in a hierarchical triple, the inner binary may achieve a certain velocity if the outer star explodes as a supernova, but an analysis of such a system has yet to be made, which is the motivation for this paper.

In this paper, we investigate the velocity distribution of remnant binaries after a tertiary explodes as a core-collapse supernova in a triple system, in order to provide constraints on the velocity distributions of high velocity runaway binaries (HVRBs), which other authors may take advantage of to probe the hypothesis that such binaries can have sufficiently high velocities consistent with the narrative that some Gaia HVS candidates could be binaries in disguise. Our methods and results are presented in the next section, while a discussion of the implications of our work is conducted in the last.

\section{Simulations and Results}

When a supernova explodes in a multiple stellar system, the velocity of the surrounding matter is modified in three ways \citep{1998A&A...330.1047T}. (i) The star that exploded as a supernova now has most (core-collapse) or all (thermonuclear) of its mass dispersed, and can no longer hold the surrounding material in orbit around itself. The original orbital velocity of the surrounding matter therefore now leads it to move away from its original position. (ii) In the case of a core-collapse supernova, the supernova may leave behind its core in the form of a neutron star, which can have a very high velocity due to the ``kick" imparted upon it by the host supernova, which interacts with the remnant system. (iii) The material driven out by the supernova shock and the shock itself may interact with the surrounding matter. For the rest of this paper, we consider only the influence of orbital velocity and the presence of the remnant neutron star, while omitting the effects of neutron star kicks and supernova shocks - we shall later see that the effects of this omission are easily accounted for.

The following work is divided into two parts. In the first part, we establish our hierarchical triple model, and introduce a set of Monte Carlo simulations, from which we generate a sample of hierarchical triples with small inner binary separations, and calculate their initial orbital velocity distribution. In the second part, we calculate the final HVRB velocity distribution generated from that sample.

\subsection{Initial Orbital Velocity Distribution}

In a hierarchical triple system consisting of three main sequence (MS) stars of masses $m_{\rm 1}$, $m_{\rm 2}$ and $m_{\rm 3}$, where the inner binary consists of $m_{\rm 1}$ and $m_{\rm 2}$, the outer tertiary $m_{\rm 3}$ would eventually evolve into a core-collapse supernova if it has a suitable initial mass, and if mass tranfer between the bodies can be disregarded. When this happens, most of the mass of $m_{\rm 3}$ dissipates, leaving behind a neutron star roughly a tenth of the original MS star's mass (see \citealt{2003ApJ...591..288H,2017ApJ...850L..19M}). Note that we later model this remnant neutron star as a point mass exactly a tenth of the original MS star's mass.

Since the velocity of the final remnant binary is heavily dependent upon its outer orbital velocity, we require a close triple system with a small outer binary separation $a_{\rm 2}$ to produce a HVRB. For a given inner binary separation $a_{\rm 1}$, the smallest $a_{\rm 2}$ that does not lead to an unstable system is

\begin{equation}
a_{\rm 2,crit}=2.8a_{\rm 1}\left[\left(1+q_{\rm 2}\right)\frac{1+e_{\rm 2}}{\sqrt{1-e_{\rm 2}}}\right]^{0.4}\left(1-\frac{0.3i}{\pi}\right)
\label{YMA}
\end{equation}

\noindent where $q_{\rm 2}=m_{\rm 3}/\left(m_{\rm 1}+m_{\rm 2}\right)$ is the mass ratio of the outer orbit, $e_{\rm 2}$ is the eccentricity of the outer orbit, and $i$ is the inclination angle between the inner and outer orbits in radians (\citealt{2001MNRAS.321..398M}, \citealt{2018MNRAS.474...20H}, see also \citealt{1995ApJ...455..640E}). As the minimum value of $a_{\rm 2}$ is proportional to $a_{\rm 1}$, we consider only systems where $a_{\rm 1}$ is between $3R_{\odot}$ and $9R_{\odot}$. For simplicity, we assume $e_{\rm 2}=0$, i.e. circular outer orbits. We further simplify the system by omitting tidal effects, allowing for all three stars to be approximated as point masses.

We then proceed to use a Monte Carlo method to generate a sample of such close hierarchical triple systems as follows. We generate each of the three masses, $m_{\rm 1}$, $m_{\rm 2}$, and $m_{\rm 3}$, by drawing them randomly from a Salpeter IMF \citep{1955ApJ...121..161S}, the low-mass cutoff of which is set at $0.5M_{\odot}$. We also generate inner and outer orbital periods $P_{\rm 1}$ and $P_{\rm 2}$, such that $\log{P}$ follows a flat distribution in both cases. It should be noted that this is directly analogous to distributions that other authors have used \citep[e.g.][]{2019MNRAS.482.3656R}. The ranges of $\log{P}$ from which the orbital periods are generated are $\log{[P_{\rm 1}/\rm (1yr)]}=-3$ to -2 (hence generating an outer orbital period between 0.3 days and 3 days for the inner orbit) and $\log{[P_{\rm 2}/\rm (1yr)]}=-3$ to 2 (hence generating an outer orbital period between 0.3 days and 100 years for the outer orbit). The rationale behind the former is that all values of $a_{\rm 1}$ derived from this range lies within the $3R_{\odot}$ to $9R_{\odot}$ range that we study. The inclination angle $i$ is assumed to follow a flat distribution from $0$ to $\pi$, for want of a better approximation. All systems are required to match the following criteria:
\begin{itemize}
  \item $14M_{\odot}<m_{\rm 3}<20M_{\odot}$. We need $m_{\rm 3}$ to evolve into a core-collapse supernova, for which this mass range is ideal.
  \item The zero age main sequence (ZAMS) and terminal age main sequence (TAMS) radii of both $m_{\rm 1}$ and $m_{\rm 2}$ must be smaller than their Roche Lobes within the inner binary. If this criterion is violated, it is possible that the inner binary components may fill their Roche Lobes during the MS phase, leading to complications. The stellar radii were calculated via \cite{1991Ap&SS.181..313D}, while the Roche Lobe sizes were calculated via \cite{1983ApJ...268..368E}.
  \item $3R_{\odot}<a_{\rm 1}<9R_{\odot}$, because this is the range we study, as noted above.
  \item The value of $a_{\rm 2}$, calculated via $m_{\rm 1}$, $m_{\rm 2}$, $m_{\rm 3}$, $P_{\rm 1}$ and $P_{\rm 2}$, must be larger than or equal to the critical value $a_{\rm 2,crit}$ given by Eq. \ref{YMA}.
\end{itemize}
Also implicit in these requirements is that $m_{\rm 3}$ must be much larger than $m_{\rm 1}+m_{\rm 2}$, so that it will explode as a supernova long before either $m_{\rm 1}$ or  $m_{\rm 2}$ can evolve. This is ascertained by the first two criteria. If any of these criteria are violated, the system is discarded and regenerated.

Thus, we generate $10^5$ hierarchical triple systems. Since we have already assumed that $e_{\rm 2}=0$, calculation of the orbital velocity distribution of the inner binary centres of mass is trivial:
\begin{equation}
v=\left(\frac{m_{\rm 3}}{m_{\rm 1}+m_{\rm 2}+m_{\rm 3}}\right)\sqrt{\frac{\mu}{a_{\rm 2}}},
\label{velocity}
\end{equation}
\noindent where ${\mu}=G(m_{\rm 1}+m_{\rm 2}+m_{\rm 3})$.

\begin{figure}
\includegraphics[scale=0.37, angle=0, trim= 3cm 2cm 0cm 0cm]{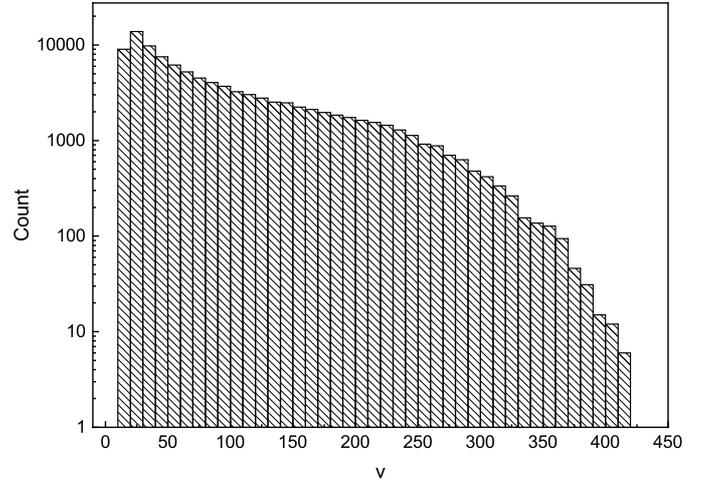}
\caption{Initial orbtial velocity distribution of inner binaries in our Monte Carlo sample. The masses of these systems are drawn from a Salpeter IMF, while the orbital separations are determined by assuming a flat $\log{P}$ distribution for the orbital periods. All systems satisfy the selection criteria that the tertiary has an appropriate mass, the inner binary does not undergo Roche Lobe overflow, the inner binary separation is between $3R_{\odot}$ and $9R_{\odot}$, and the triple system is dynamically stable when the components are at initial masses. The final orbital velocities are calculated using Eq. \ref{velocity}. The velocity distribution undergoes a sharp cutoff at around $v=400$ km/s. It should be noted that this distribution can be used as a proxy for the HVRB velocity once neutron star kicks are considered (see discussion for details). \label{Fig1}}
\end{figure}

The distribution of these initial orbital velocities is plotted in Fig. \ref{Fig1}, where it can be seen that the velocity undergoes a sharp cutoff at around $v=400$ km/s.

\subsection{Final Runaway Binary Velocity Distribution}

As the components of the stellar triple evolve, $m_{\rm 3}$ undergoes core-collapse long before $m_{\rm 1}$ or $m_{\rm 2}$ leave the MS. The supernova event dissipates most of the mass of $m_{\rm 3}$, leaving only a neutron star in its wake. The sudden mass loss may or may not unbind the outer orbit of the triple system - if it does, and if the resulting velocity of the inner binary after escaping the gravitational potential of the remnant neutron star of $m_{\rm 3}$ is significant, a HVRB composed of the inner double MS binary is born.

To simulate the effects of a core-collapse supernova in $m_{\rm 3}$, we set its mass to exactly one tenth of its initial amount ($m_{\rm NS}=m_{\rm 3}/10$), while retaining its original orbital velocity (no kick). The mass loss removes from the system a quantity of momentum equal to
\begin{equation}
\left(m_{\rm 3}-m_{\rm NS}\right)\left(\frac{m_{\rm 1}+m_{\rm 2}}{m_{\rm 1}+m_{\rm 2}+m_{\rm 3}}\right)\sqrt{\frac{\mu}{a_{\rm 2}}},
\label{momentum}
\end{equation}
\noindent and the remnant system's centre of mass consequently attains a velocity of 
\begin{equation}
v_{\rm COM}=\left(\frac{m_{\rm 3}-m_{\rm NS}}{m_{\rm 1}+m_{\rm 2}+m_{\rm NS}}\right)\left(\frac{m_{\rm 1}+m_{\rm 2}}{m_{\rm 1}+m_{\rm 2}+m_{\rm 3}}\right)\sqrt{\frac{\mu}{a_{\rm 2}}}
\label{vcom}
\end{equation}
\noindent relative to the centre of mass of the original progenitor system (prior to the supernova explosion).

In the reference frame where the remnant system's centre of mass is stationary, the velocity of the inner binary is
\begin{equation}
v_{\rm inner}=\left(\frac{m_{\rm NS}}{m_{\rm 1}+m_{\rm 2}+m_{\rm NS}}\right)v_{\rm rel},
\label{vinner}
\end{equation}
\noindent where $v_{\rm rel}$ is the relative velocity between the inner binary and the remnant neutron star (Jacobian coordinates). After the remnant binary escapes to infinity, $v_{\rm rel}$ can be calculated via

\begin{equation}
\frac{1}{2}v_{\rm rel}^{2}=\frac{1}{2}v_{\rm init}^{2}-\frac{G(m_{\rm 1}+m_{\rm 2}+m_{\rm NS})}{a_{\rm 2}},
\label{vrel}
\end{equation}
\noindent where $v_{\rm init}=\sqrt{{\mu}/a_{\rm 2}}$ is the original relative velocity at the time of the explosion, which is equal to the relative velocity prior to the explosion. Needless to say, if the right-hand side of Eq. \ref{vrel} is smaller than zero, the system remains bound, the inner binary cannot escape to infinity from the remnant neutron star, and no HVRB is born. However, this is not the case for any of our systems.

\begin{figure}
\includegraphics[scale=0.37, angle=0, trim= 3cm 2cm 0cm 0cm]{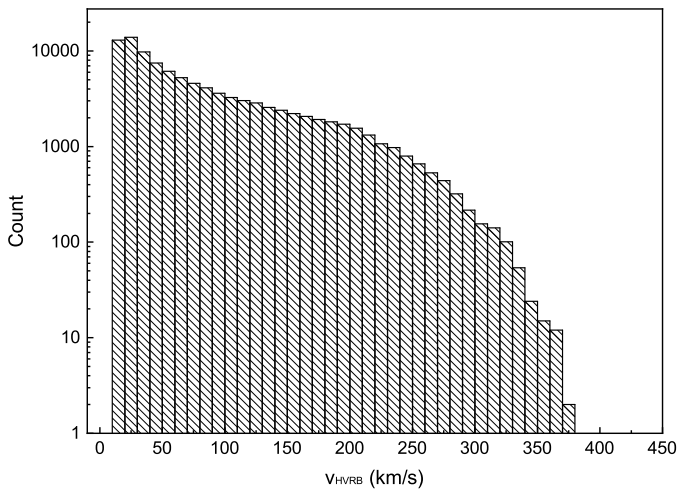}
\caption{Projected ejection velocity distribution of a population of HVRBs generated from our Monte Carlo sample of triple systems, the initial orbital velocities of which are plotted in Fig. \ref{Fig1}. The velocity values are calculated via Eqs. \ref{vcom} to \ref{vhvrb}. The velocity distribution undergoes a sharp cutoff at around $v=350$ km/s. This velocity is usually insufficient to escape the Galactic potential, but once combined with Galactic rotation or Large Magellanic Cloud effects, could lead to enhanced velocities which can. \label{Fig2}}
\end{figure}

The final velocity of the inner binary after it has escaped the remnant neutron star in the reference frame of the centre of mass of the original progenitor system is taken to be the final HVRB velocity, which is equal to:
\begin{equation}
v_{\rm HVRB}=v_{\rm inner}+v_{\rm COM}.
\label{vhvrb}
\end{equation}
\noindent This is calculated for all the triple systems in our Monte-Carlo sample, and the distribution of the final HVRB velocities is plotted in Fig. \ref{Fig2}, where a velocity cutoff at around $v=350$ km/s can be seen. 

\section{Discussion}

In this paper, we have studied the production of runaway binaries originating from supernova explosions in the tertiaries of close hierarchical triple systems. Throughout our investigation, we made several implicit assumptions, which we bring to the attention of the reader here. 

Before the tertiary ($m_{\rm 3}$) becomes a supernova, it will unavoidably become a giant first, and fill its Roche Lobe. Throughout our simulations, we disregard this process. However, it should be noted that previous work on binary supernova HVSs have allowed the Roche Lobe overflow of the exploding star, opting instead to avoid the formation of a common envelope \citep[e.g.][]{2015MNRAS.448L...6T}. Unless the overflow material can cause the inner binary to merge in the systems we simulate (which is intuitively unlikely), disregarding such Roche Lobe overflow becomes indistinguishable from the protocols hitherto adopted.

Perhaps more relevant are the issues of Lidov-Kozai resonance (\citealt{1962AJ.....67..591K}, see also \citealt{2016ARA&A..54..441N}) and tertiary tides (\citealt{2018MNRAS.479.3604G}, see also \citealt{2013MNRAS.429.2425F}), which can alter the dynamical evolution of the inner binary. Lidov-Kozai resonance, acting on triple systems with high relative orbital inclination, would inject eccentricity into the inner binary, causing it to undergo Roche Lobe overflow more easily. As for the latter, since it is inevitable that the tertiary will fill its Roche Lobe prior to its supernova explosion, tertiary tides will serve to harden the inner binary before the HVRB receives its velocity. This can affect the evolution of the inner binary, leading to complications. All these effects will presumably be more of a problem for inner binaries that are more massive (and hence have components with larger radii), which will have lower final velocities, and as such will be less relevant to the high-velocity end of our results.

The two aspects of supernova explosions which we chose to leave out of our calculations, namely the neutron star kicks and shock waves from supernovae, also deserve attention. After all, judging by the results of \cite{2015MNRAS.448L...6T}, these factors can result in a final ejection velocity twice that of the initial orbital velocity for HVSs. However, this is only the case for extreme neutron star kicks of $\sim$1000 km/s. For more common kick velocities of less than 500 km/s \citep{2016MNRAS.456.4089B}, the final HVS velocities do not exceed their initial orbital velocities, even with perfect kick alignment. Therefore, our initial orbital velocity results in Fig. \ref{Fig1} can be taken to be an estimate of how fast our HVRBs could have been for systems with normal kicks, assuming that the direction of the kick is always perfect for accelerating the companion binary, and that the neutron star never disrupts the binary as a result.

Last but not least, it should be noted that the orbital periods used in our study are taken from a flat $\log{P}$ distribution. This practice is common in studies of low-mass binaries, since the orbital period distributions of such binaries do not deviate far from a flat distribution in the local universe \citep[e.g.][]{1991A&A...248..485D}. However, The orbital period distribution of hierarchical triples has not been studied so extensively, especially for hierarchical triples with massive tertiaries, and this assumption may consequently under- or overestimate the amount of orbits at a certain orbital period, resulting in a bias. To demonstrate the extent of such a potential bias, we repeat our simulations, this time assuming a fixed value for $a_{\rm 1}$ of $6R_{\odot}$, thereby eliminating all the highest-velocity HVRBs emanating from systems with inner orbital separations in between 3 and $6R_{\odot}$ and their corresponding short orbital periods. The results are almost indentical to our original results, except that the high-velocity cutoff occurs at roughly 350km/s and 300km/s in the counterparts of Figs. \ref{Fig1} and \ref{Fig2}, respectively. We also try a lognormal distribution for logP with a mean of log(P/days)=5.03 and a standard deviation of 2.28 \citep{2010ApJS..190....1R}, which is an alternative logP distribution adopted by many authors in recent years \citep[e.g.][]{2017A&A...602A..16T}. Substituting the flat logP distribution with this more complicated distribution, while keeping our other methods intact (including demanding that $\log{[P_{\rm 1}/\rm (1yr)]}=-3$ to -2 and $\log{[P_{\rm 2}/\rm (1yr)]}=-3$ to 2), yields similar results to our original analysis, with the cutoff velocities undergoing only a 20 km/s shift towards the low velocity end.

With these issues in mind, what are the implications of this work? First of all, it should be pointed out that the systems herein studied, namely, HVRBs generated by core-collapse explosions of massive tertiaries in close hierarchical triples, are likely to form the bulk of the high-velocity HVRB population generated far from the Galactic centre. As previously mentioned, HVRBs generated by inner binary supernova explosions require SMBH tertiaries to form HVRBs \citep{2019MNRAS.484.1506L}, and even in that case are incapable of forming MS-MS binaries like our model could, while dynamical ejection from globular clusters result in HVRBs of far lower velocities \citep{2012ApJ...751..133P}. Tertiaries exploding as Type Ia supernovae are infeasible, since there is no evidence of CO white dwarf tertiaries being able to accrete matter from an inner binary. Triple systems in central configurations are highly unstable for triple systems of comparable mass, and thus cannot exist long enough to form a HVRB. Finally, while it cannot be ruled out that chaotic triples can survive long enough for a stellar component to evolve beyond the MS phase \citep{1967AJ.....72.1187S}, such occurrences are rare, and we do not expect these formation scenarios to affect the real-world HVRB velocity distribution to a high degree of significance.

Next would be the issue of how well our sample reflects real-world HVRB progenitors. The fraction of stellar systems in the real world that can be represented by the sample we generated using our Monte Carlo simulations can be shown to be
\begin{equation}
f=3{\times}10^{-8}\left(\frac{f_{\rm tr}}{10\%}\right)\left(\frac{f_{\rm i}}{0.3\%}\right)\left(\frac{f_{\rm M}}{0.01\%}\right)
\label{drake}
\end{equation}
\noindent where $f_{\rm tr}$ is the fraction of stellar systems that are hierarchical triples, $f_{\rm i}$ is the fraction of hierarchical triples with inner binary separations within the range of our simulations, and $f_{\rm M}$ is the fraction of hierarchical triples that have tertiaries in the mass range 14 to 20$M_{\odot}$, allowing them to undergo core collapse. $f_{\rm tr}$ has been shown to be on the order of $10\%$ by many authors in the field \citep[e.g.][]{2010ApJS..190....1R,2014ApJ...794..122M}, whereas $f_{\rm M}$ is roughly $0.01\%$ according to most mainstream IMFs. It should be noted here that tertiaries in hierarchical triples tend to be statistically less massive than the inner binary components in general, but we do not expect this to affect our conclusions below. As for $f_{\rm i}$, a brief glance at the contents of the Multiple Star Catalogue (\citealt{1997A&AS..124...75T,2010yCat..73890925T}, see also \citealt{2018ApJS..235....6T}) reveals that roughly 0.3\% of all hierarchical multiple systems in that sample have inner binary separations between $3R_{\odot}$ and $9R_{\odot}$, which is the range we study. However, we caution the reader that the MSC is prone to certain selection effects, the correction of which is beyond the scope of our work. If it can be assumed that $f=3{\times}10^{-8}$, then the Galaxy, with its $10^{11}$ stars, ought to yield roughly $3{\times}10^{3}$ triple systems analogous to those in our sample. This is roughly 1/30 the sample size of our simulated triple systems.

In other words, we conclude that it should be possible that a HVRB of roughly 350 km/s can be found in the Milky Way which was generated in the manner here studied, with greater velocities of up to about 400 km/s possible with neutron star kicks and supernova shocks, but that greater velocities relative to the centre of mass of the original triple are unlikely. These velocities are generally not high enough to be unbound from the Galaxy. However, once the Galactic rotational velocity of roughly 240 km/s \citep{2016ARA&A..54..529B} or HVRBs hailing from the LMC (378km/s relative to the Milky Way, see \citealt{2017MNRAS.469.2151B}) are considered, velocities capable of escaping the Galaxy could potentially be achieved. As such, the production channel studied in this paper ought to be considered, alongside other possibilities such as triple systems in which one component undergoes absorption by a supermassive black hole \citep{2009ApJ...698.1330P}, if one is ever found.

Lastly, we consider the effect that this conclusion may have on the velocity distribution of classical HVSs. If inner binaries in hierarchical triples could indeed be accelerated to velocities in excess of 300 km/s, one would expect that much faster HVSs would be produced if and when the more massive star of the HVRB evolves into a supernova, assuming that the velocity of the HVRB and that of the resultant HVS are aligned. No evidence of such a two-stage HVS exists. Yet while we have no concrete evidence of the existence of such objects, it should be noted that several abnormally fast B-type HVSs have been noted by \cite{2018A&A...620A..48I}. These objects, being of spectral type B, could not have attained velocities much faster than 500km/s, which they actually possess, from binary supernova kicks \citep{2015MNRAS.448L...6T}, and could not have been accelerated by the Galactic central black hole, since their reversed trajectories point nowhere near the Galactic centre. But before we get carried away and declare that these objects evolved from HVRBs, it should be noted that other more viable explanations also exist for these objects, as mentioned in \cite{2018A&A...620A..48I}, such as dynamical ejection from unstable multiple stellar sytems in which most of the other components are far more massive than the ejected component.

\section*{Acknowledgements}

This work was jointly supported by the Natural Science Foundation of China (Grant No. 11521303), and the Science and Technology Innovation Talent Programme of Yunnan Province (Grant No. 2017HC018).

%%%%%%%%%%%%%%%%%%%%%%%%%%%%%%%%%%%%%%%%%%%%%%%%%%

%%%%%%%%%%%%%%%%%%%% REFERENCES %%%%%%%%%%%%%%%%%%

% The best way to enter references is to use BibTeX:

%\bibliographystyle{mnras}
%\bibliography{example} % if your bibtex file is called example.bib

\begin{thebibliography}{99}



\bibitem[\protect\citeauthoryear{Beniamini \& Piran}{2016}]{2016MNRAS.456.4089B} Beniamini P., Piran T., 2016, MNRAS, 456, 4089 

\bibitem[\protect\citeauthoryear{Blaauw}{1961}]{1961BAN....15..265B} Blaauw A., 1961, BAN, 15, 265 

\bibitem[\protect\citeauthoryear{Bland-Hawthorn \& Gerhard}{2016}]{2016ARA&A..54..529B} Bland-Hawthorn J., Gerhard O., 2016, ARA\&A, 54, 529 

\bibitem[\protect\citeauthoryear{Boubert et al.}{2017}]{2017MNRAS.469.2151B} Boubert D., Erkal D., Evans N.~W., Izzard R.~G., 2017, MNRAS, 469, 2151 

\bibitem[\protect\citeauthoryear{Boubert et al.}{2019}]{2019MNRAS.tmp..280B} Boubert D., et al., 2019, MNRAS,  

\bibitem[\protect\citeauthoryear{Bromley et al.}{2018}]{2018ApJ...868...25B} Bromley B.~C., Kenyon S.~J., Brown W.~R., Geller M.~J., 2018, ApJ, 868, 25 

\bibitem[\protect\citeauthoryear{Demircan \& Kahraman}{1991}]{1991Ap&SS.181..313D} Demircan O., Kahraman G., 1991, Ap\&SS, 181, 313 

\bibitem[\protect\citeauthoryear{Duquennoy \& Mayor}{1991}]{1991A&A...248..485D} Duquennoy A., Mayor M., 1991, A\&A, 248, 485 

\bibitem[\protect\citeauthoryear{Eggleton}{1983}]{1983ApJ...268..368E} Eggleton P.~P., 1983, ApJ, 268, 368 

\bibitem[\protect\citeauthoryear{Eggleton \& Kiseleva}{1995}]{1995ApJ...455..640E} Eggleton P., Kiseleva L., 1995, ApJ, 455, 640 

\bibitem[\protect\citeauthoryear{Fuller et al.}{2013}]{2013MNRAS.429.2425F} Fuller J., Derekas A., Borkovits T., Huber D., Bedding T.~R., Kiss L.~L., 2013, MNRAS, 429, 2425 

\bibitem[\protect\citeauthoryear{Gaia Collaboration et al.}{2018}]{2018A&A...616A...1G} Gaia Collaboration, et al., 2018, A\&A, 616, A1 

\bibitem[\protect\citeauthoryear{Gao et al.}{2018}]{2018MNRAS.479.3604G} Gao Y., Correia A.~C.~M., Eggleton P.~P., Han Z., 2018, MNRAS, 479, 3604 

\bibitem[\protect\citeauthoryear{Gvaramadze, Gualandris, \& Portegies Zwart}{2009}]{2009MNRAS.396..570G} Gvaramadze V.~V., Gualandris A., Portegies Zwart S., 2009, MNRAS, 396, 570 

\bibitem[\protect\citeauthoryear{Hattori et al.}{2018}]{2018ApJ...866..121H} Hattori K., Valluri M., Bell E.~F., Roederer I.~U., 2018, ApJ, 866, 121 

\bibitem[\protect\citeauthoryear{He \& Petrovich}{2018}]{2018MNRAS.474...20H} He M.~Y., Petrovich C., 2018, MNRAS, 474, 20 

\bibitem[\protect\citeauthoryear{Heger et al.}{2003}]{2003ApJ...591..288H} Heger A., Fryer C.~L., Woosley S.~E., Langer N., Hartmann D.~H., 2003, ApJ, 591, 288 

\bibitem[\protect\citeauthoryear{Hills}{1988}]{1988Natur.331..687H} Hills J.~G., 1988, Natur, 331, 687 

\bibitem[\protect\citeauthoryear{Irrgang, Kreuzer, \& Heber}{2018}]{2018A&A...620A..48I} Irrgang A., Kreuzer S., Heber U., 2018, A\&A, 620, A48 

\bibitem[\protect\citeauthoryear{Kozai}{1962}]{1962AJ.....67..591K} Kozai Y., 1962, AJ, 67, 591 

\bibitem[\protect\citeauthoryear{Lu \& Naoz}{2019}]{2019MNRAS.484.1506L} Lu C.~X., Naoz S., 2019, MNRAS, 484, 1506 

\bibitem[\protect\citeauthoryear{Marchetti, Rossi, \& Brown}{2018}]{2018MNRAS.tmp.2466M} Marchetti T., Rossi E.~M., Brown A.~G.~A., 2018, MNRAS,  

\bibitem[\protect\citeauthoryear{Mardling \& Aarseth}{2001}]{2001MNRAS.321..398M} Mardling R.~A., Aarseth S.~J., 2001, MNRAS, 321, 398 

\bibitem[\protect\citeauthoryear{Margalit \& Metzger}{2017}]{2017ApJ...850L..19M} Margalit B., Metzger B.~D., 2017, ApJ, 850, L19 

\bibitem[\protect\citeauthoryear{Michaely \& Perets}{2014}]{2014ApJ...794..122M} Michaely E., Perets H.~B., 2014, ApJ, 794, 122 

\bibitem[\protect\citeauthoryear{Naoz}{2016}]{2016ARA&A..54..441N} Naoz S., 2016, ARA\&A, 54, 441 

\bibitem[\protect\citeauthoryear{Perets}{2009}]{2009ApJ...698.1330P} Perets H.~B., 2009, ApJ, 698, 1330 

\bibitem[\protect\citeauthoryear{Perets \& {\v S}ubr}{2012}]{2012ApJ...751..133P} Perets H.~B., {\v S}ubr L., 2012, ApJ, 751, 133 

\bibitem[\protect\citeauthoryear{Raghavan et al.}{2010}]{2010ApJS..190....1R} Raghavan D., et al., 2010, ApJS, 190, 1 

\bibitem[\protect\citeauthoryear{Rebassa-Mansergas et al.}{2019}]{2019MNRAS.482.3656R} Rebassa-Mansergas A., Toonen S., Korol V., Torres S., 2019, MNRAS, 482, 3656 

\bibitem[\protect\citeauthoryear{Salpeter}{1955}]{1955ApJ...121..161S} Salpeter E.~E., 1955, ApJ, 121, 161 

\bibitem[\protect\citeauthoryear{Shen et al.}{2018}]{2018ApJ...865...15S} Shen K.~J., et al., 2018, ApJ, 865, 15 

\bibitem[\protect\citeauthoryear{Szebehely \& Peters}{1967}]{1967AJ.....72.1187S} Szebehely V., Peters C.~F., 1967, AJ, 72, 1187 

\bibitem[\protect\citeauthoryear{Tauris \& Takens}{1998}]{1998A&A...330.1047T} Tauris T.~M., Takens R.~J., 1998, A\&A, 330, 1047 

\bibitem[\protect\citeauthoryear{Tauris}{2015}]{2015MNRAS.448L...6T} Tauris T.~M., 2015, MNRAS, 448, L6 

\bibitem[\protect\citeauthoryear{Tokovinin}{1997}]{1997A&AS..124...75T} Tokovinin A.~A., 1997, A\&AS, 124, 75 

\bibitem[\protect\citeauthoryear{Tokovinin}{2010}]{2010yCat..73890925T} Tokovinin A., 2010, yCat, 73890925

\bibitem[\protect\citeauthoryear{Tokovinin}{2018}]{2018ApJS..235....6T} Tokovinin A., 2018, ApJS, 235, 6 

\bibitem[\protect\citeauthoryear{Toonen et al.}{2017}]{2017A&A...602A..16T} Toonen S., Hollands M., G{\"a}nsicke B.~T., Boekholt T., 2017, A\&A, 602, A16 

\bibitem[\protect\citeauthoryear{Wang \& Han}{2009}]{2009A&A...508L..27W} Wang B., Han Z., 2009, A\&A, 508, L27 

\bibitem[\protect\citeauthoryear{Widmark, Leistedt, \& Hogg}{2018}]{2018ApJ...857..114W} Widmark A., Leistedt B., Hogg D.~W., 2018, ApJ, 857, 114 

\bibitem[\protect\citeauthoryear{Yu \& Tremaine}{2003}]{2003ApJ...599.1129Y} Yu Q., Tremaine S., 2003, ApJ, 599, 1129 



\end{thebibliography}

% Alternatively you could enter them by hand, like this:
% This method is tedious and prone to error if you have lots of references

%%%%%%%%%%%%%%%%%%%%%%%%%%%%%%%%%%%%%%%%%%%%%%%%%%

%%%%%%%%%%%%%%%%% APPENDICES %%%%%%%%%%%%%%%%%%%%%

%\appendix

%%%%%%%%%%%%%%%%%%%%%%%%%%%%%%%%%%%%%%%%%%%%%%%%%%

% Don't change these lines
\bsp	% typesetting comment
\label{lastpage}
\end{document}